\documentclass[11pt]{article}
\usepackage{tikz}
\usetikzlibrary{snakes}
\usepackage[utf8]{inputenc} 
\usepackage{textcomp} 
\usepackage{flafter}  

\usepackage{amsmath,amssymb}  
\usepackage{bm}  
\usepackage{authblk}
\usepackage{memhfixc}  
\usepackage{listings}
\numberwithin{equation}{section}
\begin{document}
\title{Solution to the one-dimensional telegrapher's equation subject to a backreaction boundary condition}
\author{Thorsten Pr\"ustel} 
\author{Martin Meier-Schellersheim} 
\affil{Laboratory of Systems Biology\\National Institute of Allergy and Infectious Diseases\\National Institutes of Health}
\maketitle
\let\oldthefootnote\thefootnote 
\renewcommand{\thefootnote}{\fnsymbol{footnote}} 
\footnotetext[1]{Email: prustelt@niaid.nih.gov, mms@niaid.nih.gov} 
\let\thefootnote\oldthefootnote 
\abstract
{We discuss solutions of the one-dimensional telegrapher's equation in the presence of boundary conditions. We revisit the case of a radiation boundary condition and obtain an alternative expression for the already known Green's function. Furthermore, we formulate a backreaction boundary condition, which has been widely used in the context of diffusion-controlled reversible reactions, for a one-dimensional telegrapher's equation and derive the corresponding Green's function. 
}
\section{Introduction}
Although the diffusion equation \cite{einstein1956investigations, carslaw1986conduction}
\begin{equation}\label{diffusionEq}
\frac{\partial p(x, t)}{\partial t} = D\frac{\partial^{2} p(x, t)}{\partial x^{2}} 
\end{equation}
can be successfully applied to a broad range of different phenomena, it suffers from a pathological defect: The implied signal propagation speed is infinite in the sense
 that any perturbation $\delta p(x, t)$ localized at one point in space instantaneously influences $p(x, t)$ at any other point. Still, the diffusion equation provides an excellent approximation, as long as the time scale on which inertial effects must not be neglected is small compared to the relevant time scale of the considered problem. This is not always the case, though. For instance, the movement of animals and biological cells proceeds with a rather well-defined and finite velocity \cite{Codling2008} and the diffusion equation approximation becomes invalid.
   
The telegrapher's equation \cite{Goldstein1951, MorseFeshbach, Weiss2002, Masoliver1996} may be considered as a generalization of the diffusion equation that overcomes the described deficiencies. Being a hyperbolic differential equation, it exhibits 
wave-like features on short time scales while diffusion-like features dominate on longer time scales. In particular, the signal propagation speed is bounded and well-defined. 
Like the diffusion equation, the telegrapher's equation may be viewed from a macroscopic, phenomenological and, at least in one dimension, from a microscopic point of view. 

The phenomenological approach starts with Fick's first law that asserts that the probability flux is linearly related to the gradient of the probability density 
\begin{equation}\label{Fick}
j(x,t) = -D\frac{\partial p(x,t)}{\partial x}.
\end{equation}
Combining Eq.~\eqref{Fick} with the conservation law
\begin{equation}\label{conservationEq}
\frac{\partial p}{\partial t} + \frac{\partial j}{\partial x} = 0 
\end{equation}
immediately yields the diffusion equation. Consequently, Ficks' first law and the telegrapher's equation are not compatible with each other and one has to seek a modification of Eq.~\eqref{Fick}. A suitable generalization was suggested by Cattaneo in an attempt to replace Fourier's law in the theory of heat waves. In the present context Cattaneo's equation \cite{Joseph1989} reads as
\begin{equation}\label{cattaneo}
T\frac{\partial j}{\partial t} = -\bigg(D\frac{\partial p}{\partial x} +j\bigg),
\end{equation} 
where $T$ sets  the relaxation time scale. Obviously, for $T\rightarrow 0$ one recovers Eq.~\eqref{Fick}. We note for later use that
\begin{equation}\label{defJ}
j(x, t) = -c^{2}\int^{t}_{0}e^{-(t-\tau)/T}\frac{\partial p(x,\tau)}{\partial x}\,d\tau
\end{equation}
satisfies Eq.~\eqref{cattaneo}, given that one identifies the diffusion constant with the speed $c$ and the relaxation time $T$ according to
$D=c^{2}T$.
Cattaneo's equation Eq.~\eqref{cattaneo} combined with the conservation equation Eq.~\eqref{conservationEq} leads to the telegrapher's equation
\begin{equation}\label{telegraphersEq}
\frac{\partial^{2}p}{\partial t^{2}}+\frac{1}{T}\frac{\partial p}{\partial t} = c^{2} \frac{\partial^{2}p}{\partial x^{2}}.
\end{equation}
Note that $T$ interpolates between the ballistic and diffusion regime in the sense that Eq.~\eqref{telegraphersEq} reduces to the wave equation in the limit
$T \rightarrow \infty$, while one arrives at the diffusion equation Eq.~\eqref{diffusionEq} in the limit $T\rightarrow 0, c \rightarrow \infty, c^{2}T\rightarrow D=\text{finite}$. 

Turning to a microscopic point of view, we recall the well-known fact that the diffusion equation can be obtained as the continuum limit of the simple uncorrelated random walk, also referred to as Brownian motion.  Similarly, at least in one dimension, the telegrapher's equation emerges from a correlated random walk that takes into account correlations in the direction of movement \cite{Fuerth1917, Taylor1922, Goldstein1951}. In this context Brownian motion appears as a contracted description of the more general correlated random walk, being valid only if the characteristic correlation time decays sufficiently fast. The persistence of the correlated walker, i.e. the tendency that each step points in the same direction as the previous one, necessitates to consider two separate probability density functions (PDF). Let $a(x, t\vert x_{0})$ denote the PDF to find the particle at $x$, given that it was at $x_{0}$ for $t=0$ and that it is moving in the 
positive $x$ direction at time $t$, and let $b(x, t\vert x_{0})$ the corresponding PDF for moving in the negative $x$ direction. It follows that the PDF $p(x, t\vert x_{0})$ 
independent of the direction in which the particle is moving at $t$ is given by the superposition
\begin{equation}
p(x, t\vert x_{0}) = a(x, t\vert x_{0}) + b(x, t\vert x_{0}).
\end{equation}
It can be shown that the time evolution of $p(x, t\vert x_{0})$ is governed by the telegrapher's equation Eq.~\eqref{telegraphersEq} \cite{Masoliver1993}.

The broad range of potential applications of the telegrapher's equation \cite{Weiss2002}, its blend of wave and diffusion-like features and relation to other 
important equations like the diffusion, wave and Dirac equation \cite{Gaveau1984}, motivate a thorough study of its solutions. 

In this paper, we are interested in solutions to Eq.~\eqref{telegraphersEq} in the presence of a single boundary \cite{Masoliver1993, Masoliver1992, Masoliver1992erratum}. Previously, a radiation boundary condition (BC), the purely absorbing and reflecting limiting cases and the corresponding Green's functions (GF) have been derived \cite{Masoliver1993}. Here, we will revisit the absorbing and radiation boundary cases.  We will obtain an alternative form of the GFs by directly inverting the Laplace transform via the Bromwich contour integral and we will discuss how the alternative forms relate to the expressions obtained earlier. Then, we will discuss and derive a backreaction BC suitable for a one-dimensional telegrapher's equation. The backreaction BC generalizes the radiation BC, which only describes irreversible reactions between the particle and boundary: Once the particle gets absorbed, it stays trapped forever. In contrast, the backreaction BC incorporates desorption or return of population at the boundary. Consequently, the backreaction BC figures prominently in the theory of reversible diffusion-influenced reactions \cite{Agmon:1990p10} and the GF of the diffusion equation subject to that BC have been obtained in one, two, and three dimensions \cite{Agmon:1984, TPMMS_2012JCP, kimShin:1999}. 
Finally, we will focus on the purely absorption-desorption case and calculate the associated GF by make use of the Bromwich contour integral again. In his way, we can derive and discuss all obtained GFs in a coherent way.
 
\section{Radiation boundary condition} 
We will start by summarizing some results obtained earlier \cite{Masoliver1993}.
Eq.~\eqref{telegraphersEq} has to be supplemented by initial conditions. A choice that can be motivated from an underlying random walk picture is
\begin{eqnarray}\label{initialCond}
p(x, t=0\vert x_{0}) &=& \delta(x-x_{0}),\\
\frac{\partial p}{\partial t}\bigg\vert_{t=0} &=& 0.
\end{eqnarray}
The solution to the telegrapher's equation that satisfy these initial conditions is known as the free-space GF \cite{Masoliver1993}
\begin{eqnarray}
p_{\text{free}}(x, t\vert x_{0}) = e^{-t/(2T)}f_{0}(t, \vert x- x_{0}\vert),
\end{eqnarray}
where 
\begin{eqnarray}
f_{0}(t, x)&=&\frac{1}{2}\delta(ct-x) + \frac{\Theta(ct-x)}{4cT}\bigg[I_{0}(u) + \frac{t}{2uT}I_{1}(u)\bigg],\\
u&=& \frac{\sqrt{c^{2}t^{2}-x^{2}}}{2cT}.
\end{eqnarray}
Here $\Theta(ct-x)$ denotes the Heaviside step function and $I_{0}, I_{1}$ refer to the modified Bessel functions of first kind \cite[Ch.~(9.6.)]{abramowitz1964handbook}. 
For later reference we note that the Laplace transform of $f_{0}(t, x)$ is 
\begin{equation}\label{transformF0}
\tilde{f}_{0}(s, x) = \frac{1}{2c}\frac{s+\frac{1}{2T}}{\sqrt{s^{2}-\frac{1}{4T^{2}}}}e^{-x/c\sqrt{s^{2}-\frac{1}{4T^{2}}}}
\end{equation}
The free-space GF can be used to obtain GFs that satisfy the telegrapher's equation in the presence of boundaries. Typically, one seeks a solution of the form
\begin{equation}\label{ansatzGF}
p(x,t\vert x_{0}) = p_{\text{free}}(x,t\vert x_{0}) + h(x,t),
\end{equation}
where $h(x,t)$ has to be a solution of the telegrapher's equation Eq.~\eqref{telegraphersEq}. Furthermore, it is chosen in such a way that Eq.~\eqref{ansatzGF} satisfy the specified BC.
Henceforth, we will assume without loss of generality that the  boundary is located at $x=0$ and that $x\geq 0, x_{0}\geq 0$. 

In Ref. \cite{Masoliver1993} it has been shown that the appropriate radiation BC is quite different to the one in the diffusion case and that it takes the form
\begin{equation}\label{radiationBC}
c\frac{\partial p}{\partial x}\bigg\vert_{x=0}=\frac{\beta}{2-\beta}\bigg(\frac{\partial p}{\partial t}+\frac{1}{T}p\bigg)\bigg\vert_{x=0}.
\end{equation}
Here, $\beta$ denotes the probability that a particle reaching the boundary gets actually absorbed
\begin{equation}\label{radiationBCPDF}
a(x=0, t\vert x_{0}) = (1-\beta) b(x=0, t\vert x_{0}).
\end{equation}
Thus, $\beta=1$ corresponds to the purely absorbing limit, while $\beta =0$ corresponds to a purely reflecting boundary. 
Furthermore, seeking a solution of the form
\begin{equation}
p(x, t\vert x_{0}) = e^{-t/(2T)} P(x, t\vert x_{0}),
\end{equation}
one finds for the Laplace transform of $P(x, t\vert x_{0})$ \cite{Masoliver1993}
\begin{equation}\label{transformP}
\tilde{P}(x, s\vert x_{0}) = \frac{s + \frac{1}{2T}}{2c\rho}\bigg[e^{-\rho\vert x- x_{0}\vert/c} - \Omega e^{-\rho(x + x_{0})/c}\bigg],
\end{equation}
where
\begin{eqnarray}
\rho &=& \sqrt{s^{2} - \frac{1}{4T^{2}}}, \label{defRho}\\
\Omega &=& \frac{\beta (s + \frac{1}{2T}) - (2-\beta)\rho}{\beta (s + \frac{1}{2T}) + (2-\beta)\rho}. \label{defOmega}
\end{eqnarray}
It is easy to show that for purely absorbing and reflecting BCs the expression for $\Omega$ reduces to
\begin{eqnarray}
\Omega &=& 2T(s-\rho), \,\,\quad \text{if}\,\, \beta =1,\\
\Omega &=& 1, \quad\qquad\qquad \text{if}\,\, \beta =0,
\end{eqnarray}
respectively.
Taking into account these identities, Eqs.~\eqref{transformF0}, \eqref{transformP} and elementary properties of the Laplace transform, one 
arrives at the following form for the GFs corresponding to purely absorbing and reflecting BC
\begin{eqnarray}
p_{\text{abs}}(x, t\vert x_{0}) &=& e^{-t/2T}\big[f_{0}(t, \vert x-x_{0}\vert) - f_{1}(t, x+x_{0})\big], \label{pAbs}\\
p_{\text{ref}}(x, t\vert x_{0}) &=& e^{-t/2T}\big[f_{0}(t, \vert x-x_{0}\vert) + f_{0}(t, x+x_{0})\big], 
\end{eqnarray}
where the function $f_{1}$ is given by
\begin{eqnarray}\label{defF1}
f_{1}(t,x)&=&\frac{\Theta(ct-x)}{8cT}\bigg[I_{0}(u) + 2 \bigg(\frac{ct-x}{ct+x}\bigg)^{1/2}I_{1}(u) +\nonumber\\
&+& \bigg(\frac{ct-x}{ct+x}\bigg)I_{2}(u)\bigg]. 
\end{eqnarray}
Finally, to obtain the inverse Laplace transform in the case of a radiation BC one can proceed by expanding the denominator in a power series to find  
the following expression for the GF satisfying the BC Eq.~\eqref{radiationBC} \cite{Masoliver1993}
\begin{eqnarray}\label{radGF}
p_{\text{rad}}(x, t\vert x_{0}) &=& \beta p_{\text{abs}}(x, t\vert x_{0}) + (1-\beta)p_{\text{ref}}(x, t\vert x_{0})+\nonumber\\
&-&\frac{1}{8cT}\beta(1-\beta)e^{-t/2T}\sum^{\infty}_{n=0}c_{n}g_{n}(t, x+x_{0}),
\end{eqnarray}
The coefficients are defined by
\begin{eqnarray}\label{GF}
c_{0}&=&1, \nonumber\\
c_{1}&=&4-\beta, \nonumber\\
c_{2}&=&(2-\beta)(3-\beta)+1, \nonumber\\
c_{n}&=&(2-\beta)^{3}(1-\beta)^{n-3}, \, \,\,\, n \geq 3,\nonumber\\
g_{n}(t,x)&=&\Theta(ct-x)\bigg(\frac{ct-x}{ct+x}\bigg)^{n/2}I_{n}(u). 
\end{eqnarray}
\subsection{Alternative expression for the Green's function}
In what follows, we will take a different route to derive alternative expressions for the GFs subject to purely absorbing and radiation BC. By making use of the Bromwich contour integral, we will derive an alternative form of the GFs as an important preparatory step. In fact, by employing for all discussed BC the same technique, hardly any extra work will be necessary to establish the expression for the GF that satisfies the backreaction BC. 

\begin{figure}
\includegraphics[scale=0.4]{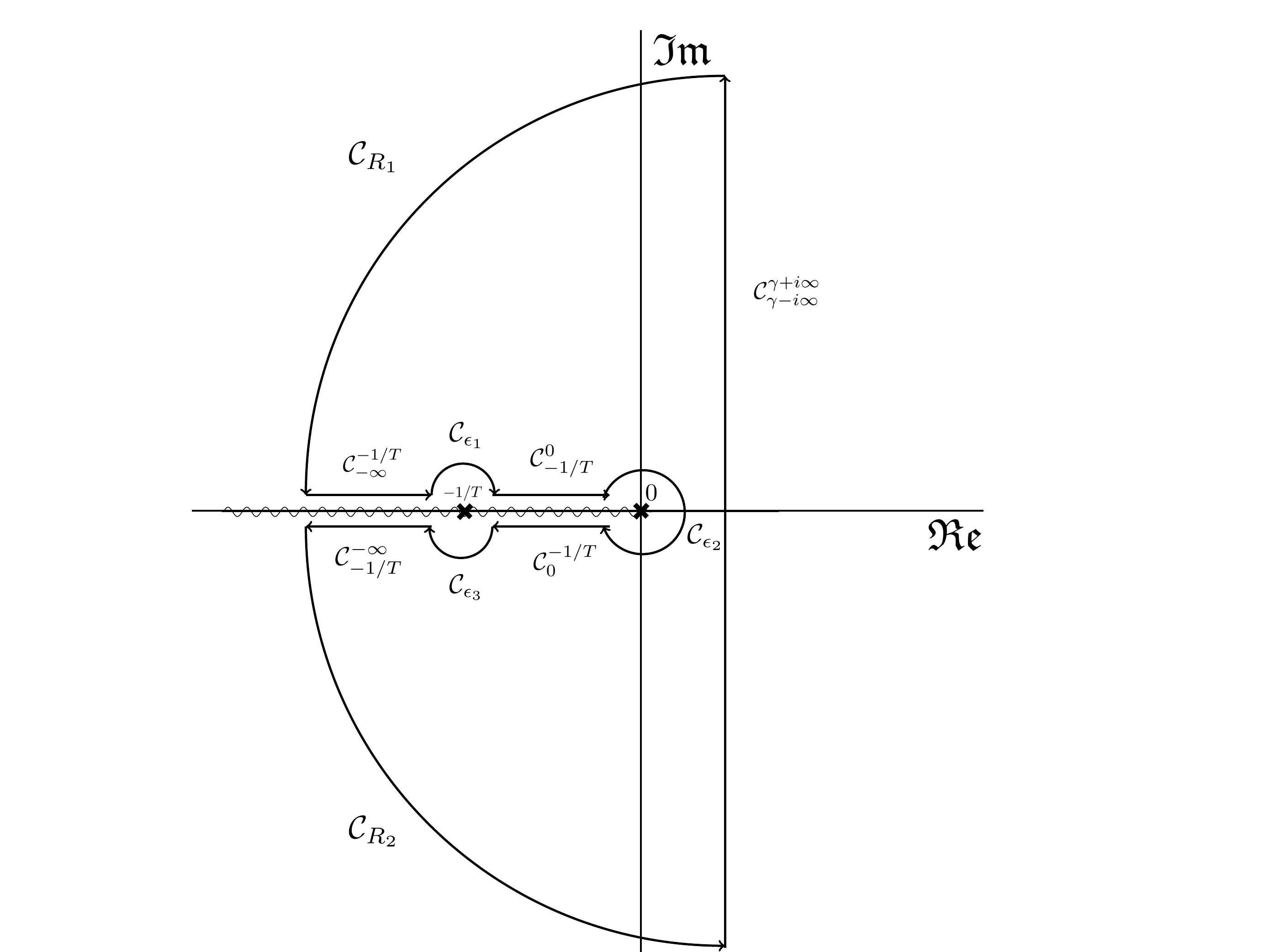}
\caption{Integration contour used  in Eq.~(\ref{Bromwich}).}\label{fig:contour}
\end{figure}

Focusing first on the purely absorbing case, we start again from the Laplace domain and write
\begin{equation}
\tilde{p}_{\text{abs}}(x,s \vert x_{0}) = \tilde{p}_{\text{free}}(x,s \vert x_{0}) + \tilde{h}_{\text{abs}}(x,s).
\end{equation}
Now, our strategy will be to invert $\tilde{h}_{\text{abs}}$ directly by using the Bromwich contour integral. Henceforth, to streamline the notation, we define
\begin{eqnarray}
a &=& T^{-1},\nonumber\\
k &=& \frac{x+x_{0}}{c}.\nonumber
\end{eqnarray}
Then, taking into account Eqs.~\eqref{transformP}, \eqref{defRho} and \eqref{defOmega}, the Bromwich contour integral takes the form
\begin{eqnarray}\label{Bromwich}
&&h_{\text{abs}}(x,t \vert x_{0}) =\frac{1}{2\pi i}\int^{\gamma+i\infty}_{\gamma-i\infty}e^{st}\tilde{h}_{\text{abs}}(x,s)ds=\nonumber\\
 &&-\frac{1}{4\pi ic}\int^{\gamma+i\infty}_{\gamma-i\infty}e^{ts - k\sqrt{s(s+a)}}\frac{s+a}{\sqrt{s(s+a)}}\frac{\sqrt{s+a}-\sqrt{s}}{\sqrt{s+a} + \sqrt{s}}ds.\quad
\end{eqnarray}
To proceed we have to discuss the analytical structure of the integrand, see Fig.~\ref{fig:contour}. First, we note that it has two branch points, at $s = -1/T$ and $s = 0$, necessitating the specification of two branch cuts. We choose both branch cuts along the negative real axis, leading to a potentially discontinuous integrand in the region $s  <  0$. Therefore, we employ the keyhole contour depicted in Fig.~\ref{fig:contour}. It turns out that the branch cuts' effects cancel each other along the overlapping region $s < -1/T$, i.e. along $\mathcal{C}^{-1/T}_{-\infty}$, $\mathcal{C}^{-\infty}_{-1/T}$, and the integrand is only discontinuous for $-1/T <  s  <  0$.  Therefore, the sole contributions coming from the integration on each sides of the branch cut are due to the integrals along $\mathcal{C}^{0}_{-1/T}$ and $\mathcal{C}^{-1/T}_{0}$.
Along $\mathcal{C}^{0}_{-1/T}$ we choose
$s = re^{i\pi}$ and it follows $\sqrt{s+a}=\sqrt{a-r}$ and $\sqrt{s} = i\sqrt{r}$. We get
\begin{equation}\label{intAlongC1}
\int_{\mathcal{C}^{0}_{-1/T}} e^{st}\tilde{h}_{\text{abs}}(x,s) ds = \int^{a}_{0}\frac{e^{-tr - ik\sqrt{r(a-r)}}}{i\sqrt{r(a-r)}}(a-r)\frac{\sqrt{a-r}-i\sqrt{r}}{\sqrt{a-r} + i\sqrt{r}}dr.
\end{equation}
Similarly, to evaluate the integral along $\mathcal{C}^{-1/T}_{0}$ , we choose $s=r e^{-i\pi}$ and note that the ensuing integral is the negative complex conjugate of the integral given in Eq.~\eqref{intAlongC1} 
\begin{equation}\label{complexC}
\int_{\mathcal{C}^{0}_{-1/T}} e^{st}\tilde{h}_{\text{abs}}(x,s) ds = - \bigg(\int_{\mathcal{C}^{-1/T}_{0}} e^{st}\tilde{h}_{\text{abs}}(x,s) ds\bigg)^{\ast},
\end{equation}
where $\ast$ denotes complex conjugation. Furthermore, the integrand has no poles neither inside nor on the contour and the contributions from the small circles $\mathcal{C}_{\epsilon_{2}}, \mathcal{C}_{\epsilon_{1}}, \mathcal{C}_{\epsilon_{3}}$ around the origin and $-1/T$, respectively, vanish.
Thus, using Eqs.~\eqref{complexC},~\eqref{intAlongC1},~\eqref{Bromwich} and making the substitutions $r\rightarrow \varphi + 1/2a,\, \varphi \rightarrow 1/2a\xi$, we finally arrive at
\begin{eqnarray}\label{finalHabs}
&&h_{\text{abs}}(x,t ) = \frac{a\Theta(ct-k)}{4\pi c}e^{-1/2at}\int^{1}_{-1}e^{-1/2at\xi}(1-\xi)\times\nonumber \\
&&\bigg[\frac{\cos(1/2ka\sqrt{1-\xi^{2}})}{\sqrt{1-\xi^{2}}}\xi + \sin(1/2ka\sqrt{1-\xi^{2}})\bigg]d\xi.
\end{eqnarray}

The question arises how the found expression relates to the expression obtained in Ref.~\cite{Masoliver1993}, cp. also Eqs.~\eqref{pAbs} and \eqref{defF1}. Actually, one can explicitly show that Eq.~\eqref{finalHabs} can be cast in an integral free form. To this end we first note the identity
\begin{equation}\label{integralBessel}
 \frac{1}{\pi}\int^{1}_{-1}e^{-1/2 a t \xi} \frac{\cos(1/2 a k \sqrt{1 - \xi^{2}})}{\sqrt{1-\xi^{2}}}d\xi = I_{0}(1/2 a\sqrt{t^{2}-k^{2}}). 
\end{equation}   
Then, it follows that the terms involving the cosine in Eq.~\eqref{finalHabs} can be rewritten by taking the first and second order derivative of the expression on the rhs of Eq.~\eqref{integralBessel} with respect to $t$.
Also, upon integration by parts we can bring the term involving the sinus in a form that can be expressed analogously.

The general case of a radiation boundary conditions can now be treated in a similar way. Again, the part of the GF that takes into account the BC can be expressed by a Bromwich integral  
\begin{eqnarray}\label{BromwichRad}
h_{\text{rad}}(x,t) =-\frac{1}{4\pi ic}\int^{\gamma+i\infty}_{\gamma-i\infty}e^{ts - k\sqrt{s(s+a)}}\times\nonumber\\
\frac{s+a}{\sqrt{s(s+a)}}\frac{(1-\eta)\sqrt{s+a}-(1+\eta)\sqrt{s}}{(1-\eta)\sqrt{s+a} + (1+\eta)\sqrt{s}}ds,
\end{eqnarray}
where $\eta = 1-\beta$. Because the analytical properties of the integrand in Eq.~\eqref{BromwichRad} are analogous to the case of purely absorbing
BC, we employ the same keyhole contour and proceed in the same way to find
\begin{eqnarray}\label{finalHrad}
&&h_{\text{rad}}(x,t) = \frac{a\Theta(ct-k)}{4\pi c}e^{-1/2at}\int^{1}_{-1}\frac{e^{-1/2at\xi}}{1+\eta^{2}+2\eta\xi}(1-\xi)\times\qquad\qquad\qquad\nonumber\\
&&\bigg[\frac{\cos(\frac{ka}{2}\sqrt{1-\xi^{2}})}{\sqrt{1-\xi^{2}}} [2\eta +(1+\eta^{2})\xi]  +[1-\eta^{2}]\sin(\frac{ka}{2}\sqrt{1-\xi^{2}})\bigg]d\xi.
\end{eqnarray} 
Here, two remarks are in order. First, as $\eta\rightarrow 0$, one recovers the expression obtained for $h_{\text{abs}}$ Eq.~\eqref{finalHabs}.
Second, we may expand the denumerator $1+\eta^{2} + 2\eta\xi$ in powers of $2\eta\xi/(1+\eta^{2})$, because $2\eta/(1+\eta^{2}) < 1$ for $\eta <1$. As explained for $h_{\text{abs}}$, the resulting integrals can be expressed as suitable time derivatives of the expression on the rhs of Eq.~\eqref{integralBessel}, and in this way one may recover the series expansion given in Eqs.~\eqref{radGF}, \eqref{GF}. 
\section{Backreaction boundary condition}
Next, we will turn to reversible interactions with the boundary: when the particle gets trapped this does not mean necessarily that it stays trapped forever. First, we will derive a backreaction BC that is suitable for the one-dimensional telegrapher's equation. The derivation will follow the line of reasoning in Ref. \cite{Masoliver1993} and it will turn out that the backreaction BC is somewhat more complicated than the corresponding BC in the case of diffusion, resembling the situation for the radiation BC.

Conventionally, the salient assumption underlying the backreaction BC is that the rate of desorption is proportional to the total absorbed population. Hence, generalizing Eq.~\eqref{radiationBCPDF},
we postulate that at the boundary the PDFs that take into account the directions are related by
\begin{equation}\label{boundaryRelationII}
a(x, t\vert x_{0})\vert_{x=0} = (1-\beta) b(x, t\vert x_{0})\vert_{x=0} + \frac{\beta\kappa}{2}[1-S(t\vert x_{0})].
\end{equation}
Here, $S(t\vert x_{0})$ denotes the survival probability, defined by
\begin{equation}\label{defS}
S(t\vert x_{0}) = \int^{\infty}_{0} p(x,t\vert x_{0})\, dx,
\end{equation}
and therefore $1-S(t\vert x_{0})$ is the total absorbed population.
The total probability density becomes
\begin{eqnarray}
p(0, t\vert x_{0}) &=& a(0, t\vert x_{0}) + b(0, t\vert x_{0}) \nonumber\\
&=& (2-\beta) b(0, t\vert x_{0}) + \frac{\beta\kappa}{2}[1-S(t\vert x_{0})].
\end{eqnarray}
Additionally, we can conclude that
\begin{equation}
a(0, t\vert x_{0}) - b(0, t\vert x_{0})= -\frac{\beta}{2-\beta}\bigg[p(0, t\vert x_{0}) - \kappa\big[1-S(t\vert x_{0})\big]\bigg].
\end{equation}
Moreover, one has the dynamic equation \cite{Masoliver1993}
\begin{equation}
c\frac{\partial p}{\partial x} = - \frac{\partial }{\partial t}(a-b) -\frac{1}{T}(a-b),
\end{equation}
which is valid in particular at the boundary $x=0$, and hence we arrive at the backreaction BC for the one-dimensional telegrapher's equation
\begin{equation}\label{backreactionBC}
c\frac{\partial p}{\partial x}\bigg\vert_{x=0} = \frac{\beta}{2-\beta}\bigg\lbrace\frac{\partial p}{\partial t}\bigg\vert_{x=0} + \kappa\frac{\partial S(t\vert x_{0})}{\partial t} + \frac{1}{T}\bigg[p\vert_{x=0} - \kappa\big(1-S(t\vert x_{0})\big)\bigg] \bigg\rbrace
\end{equation}
To shed more light on the nature of the parameter $\kappa$ it is instructive to consider the limiting case of the BC Eq.~\eqref{backreactionBC}. Taking the limit $c\rightarrow\infty$, $T\rightarrow 0$ and $\beta/(2-\beta)\rightarrow 0$ such that $c^{2}T=:D$ and $\beta/(2-\beta)c=:\kappa_{a}$ are kept finite, we recover the radiation-backreaction BC \cite{Agmon:1984} for the diffusion equation
\begin{equation}
D\frac{\partial p}{\partial x}\bigg\vert_{x=0} = \kappa_{a}p\vert_{x=0} - \kappa_{d}\big(1-S(t\vert x_{0})\big),
\end{equation}
where $\kappa_{d} := \kappa_{a} \kappa$. On the other hand, when we take the same limit as before with the exception that $\beta/(2-\beta)c\rightarrow \infty$, we arrive at the absorption-backreaction BC \cite{Agmon:1984}
\begin{equation}
p\vert_{x=0} = \kappa\big[1-S(t\vert x_{0})\big].
\end{equation}

To derive a GF of the telegrapher's equation subject to the BC specified by Eq.~\eqref{backreactionBC}, one has to find a suitable expression for the survival probability. To this end, we briefly recall how one proceeds in the case of diffusion.
From the equation of continuity Eq.~\eqref{conservationEq} and the definition of the survival probability Eq.~\eqref{defS} it follows
\begin{equation}\label{SandJ}
\frac{\partial S(t\vert x_{0})}{\partial t} = j(x,t\vert x_{0})\vert_{x=0}. 
\end{equation}
Assuming that the current vanishes at infinity and that Fick's first law Eq.~\eqref{Fick} is valid one obtains
\begin{equation}\label{expS}
S(t\vert x_{0}) = 1 - \int^{t}_{0} \frac{\partial p(x,t\vert x_{0})}{\partial x}\bigg\vert_{x=0} \,dt^{\prime}.
\end{equation}
Obviously, Eq.~\eqref{expS} cannot be used  in the context of the telegrapher's equation, because we already discussed that Fick's law has to be replaced by Cattaneo's equation Eq.~\eqref{cattaneo}. Hence, we combine Eq.~\eqref{defJ} and Eq.~\eqref{SandJ}, leading to 
\begin{equation}\label{defSGeneral}
S(t\vert x_{0}) = 1- c^{2}\int^{t}_{0}\int^{t^{\prime}}_{0}e^{-(t^{\prime}-\tau)/T}\frac{\partial p(x,\tau\vert x_{0})}{\partial x}\,d\tau \,d t^{\prime}. 
\end{equation}

The derived expression for the survival probability can be introduced in the BC Eq.~\eqref{backreactionBC}. Henceforth, we will focus on the case $\beta = 1.$ Then, in the Laplace domain the BC becomes
\begin{equation}
\bigg( c+\frac{\kappa c^{2}}{s}\bigg)\frac{\partial \tilde{p}}{\partial x}\bigg\vert_{x=0} = \bigg(s+\frac{1}{T}\bigg)\tilde{p}\bigg\vert_{x=0}. 
\end{equation}
As before, we make the ansatz
\begin{equation}\label{ansatz}
\tilde{p}(x, s\vert x_{0}) = \tilde{p}_{\text{free}}(x, s\vert x_{0}) + \tilde{h}_{\text{back}}(x, s). 
\end{equation}
The function $h_{\text{back}}(x, t)$ has to be a solution of the telegrapher's equation and hence takes in the Laplace domain the form 
\begin{equation}
\tilde{h}_{\text{back}}(x, s) = A(s) e^{-x/c\sqrt{s(s+\frac{1}{T})}}.
\end{equation}
The constant $A(s)$ must be chosen in such a way that Eq.~\eqref{ansatz} satisfies the BC. We obtain
\begin{equation}
A(s) = -\frac{1}{2c}\frac{s+a}{\sqrt{s(s+a)}}\frac{s+a - \big(1+\frac{c\kappa}{s} \big)\sqrt{s(s+a)}}{s+a + \big(1+\frac{c\kappa}{s} \big)\sqrt{s(s+a)}}
e^{-\frac{x_{0}}{c}\sqrt{s(s+a)}}.
\end{equation}
As before, we can calculate $h_{\text{back}}(x, t)$ via the Bromwich integral  
\begin{eqnarray}\label{BromwichBack}
h_{\text{back}}(x,t) =-\frac{1}{4\pi i c}\int^{\gamma+i\infty}_{\gamma-i\infty}e^{ts - k\sqrt{s(s+a)}}\times \nonumber\\
\frac{s+a}{\sqrt{s(s+a)}}\bigg[\frac{\sqrt{s+a}-\big(1+\frac{c\kappa}{s} \big)\sqrt{s}}{\sqrt{s+a} + \big(1+\frac{c\kappa}{s} \big)\sqrt{s}}\bigg]ds.
\end{eqnarray}
We can closely follow the line of reasoning presented for the case of absorbing and radiation BC. We may specify the same branch cut structure, employ the same integration contour and make the same substitutions.
Thus, we arrive at
\begin{eqnarray}\label{finalHBack}
&&h_{\text{back}}(x,t \vert x_{0}) = \frac{a\Theta(ct-k)}{4\pi c}e^{-1/2at}\int^{1}_{-1}\frac{e^{-1/2at\xi}(1-\xi)}{c^{2}\kappa^{2}+a/2(a-2c\kappa)(1+\xi)}\times\qquad\qquad\nonumber\\
&&\bigg[\frac{\cos(1/2ka\sqrt{1-\xi^{2}})}{\sqrt{1-\xi^{2}}}\Pi(\xi) + \sin(1/2ka\sqrt{1-\xi^{2}})\Gamma(\xi)\bigg]d\xi.
\end{eqnarray} 
Here, we introduced the functions
\begin{eqnarray}
\Pi(\xi) &=& \frac{1}{2} a^{2}\xi^{2} + \frac{1}{2}a(a-2c\kappa)\xi - c\kappa + c^{2}\kappa^{2},\\
\Gamma(\xi) &=& \frac{1}{2} a^{2}\xi + \frac{1}{2}a(a-2c\kappa). 
\end{eqnarray}
In the limit $\kappa\rightarrow 0$, we recover the result for the purely absorbing BC Eq.~\eqref{finalHabs}. 
Furthermore, we note that 
\begin{equation}
\bigg\vert\frac{\frac{a}{2}(a-2c\kappa)}{\frac{a}{2}(a-2c\kappa)+c^{2}\kappa^{2}}\bigg\vert < 1,
\end{equation} 
and therefore in this case also it is possible to expand the denumerator in a power expansion in
$$ \frac{\frac{a}{2}(a-2c\kappa)}{\frac{a}{2}(a-2c\kappa)+c^{2}\kappa^{2}}\xi.$$
Using Eq.~\eqref{integralBessel} this means that we can dispense with the integral representation altogether in the case of the backreaction BC also. 
\newpage
\subsection*{Acknowledgments}
This research was supported by the Intramural Research Program of the NIH, National Institute of Allergy and Infectious Diseases. 

We would like to thank Bastian R. Angermann and Frederick Klauschen for stimulating discussions.
\bibliographystyle{plain} 

\end{document}